Reduction of interface state density in SiC (0001) MOS structures by low-oxygen-partial-pressure annealing

Takuma Kobayashi[1)], Keita Tachiki[1)], Koji Ito[1)], and Tsunenobu Kimoto[1)]

[1]Department of Electronic Science and Engineering, Kyoto University, Nishikyo, Kyoto 615-8510, Japan

We report that annealing in low-oxygen-partial-pressure (low-$p_{O2}$) ambient is effective in reducing the interface state density ($D_{IT}$) at a SiC (0001)/SiO$_2$ interface near the conduction band edge ($E_C$) of SiC. The $D_{IT}$ value at $E_C - 0.2$ eV estimated by a high (1 MHz)-low method is $6.2 \times 10^{12}$ eV$^{-1}$cm$^{-2}$ in as-oxidized sample, which is reduced to $2.4 \times 10^{12}$ eV$^{-1}$cm$^{-2}$ by subsequent annealing in O$_2$ (0.001%) at 1500°C, without interface nitridation. Although annealing in pure Ar induces leakage current in the oxide, low-$p_{O2}$ annealing ($p_{O2}$ = 0.001 - 0.1 %) does not degrade the oxide dielectric property (breakdown field ~ 10.4 MVcm$^{-1}$).

Silicon carbide (SiC) is a suitable material for power device applications, owing to its superior physical properties, such as wide bandgap, high critical electric field, and high thermal conductivity.[1,2] A unique advantage of SiC over other compound semiconductors is that it can be thermally oxidized to give high-quality silicon dioxide ($SiO_2$).[2] Thus, SiC metal-oxide-semiconductor field effect transistors (MOSFETs) have attracted much attention for low-loss and fast power switches. SiC MOSFETs have, however, suffered from the low channel mobility due to the high interface state density ($D_{IT}$, ~ $10^{13}$ eV$^{-1}$ cm$^{-2}$) of SiC/$SiO_2$ systems.[2-7]

Although the physical origin of the interface states remains uncertain, several methods were found to passivate the defect levels. For instance, post-oxidation annealing in nitric oxide (NO)[8,9] or nitrous oxide ($N_2O$)[10,11] (interface nitridation) or in a gas mixture of phosphoryl chloride ($POCl_3$), oxygen ($O_2$), and nitrogen ($N_2$)[12] ($POCl_3$ annealing) is effective in reducing the $D_{IT}$ near the conduction band edge ($E_C$) of SiC. However, since these methods rely on the incorporation of foreign atoms (nitrogen (N) or phosphorus (P)), generation of extrinsic defects at the interface and in the oxide have been pointed out as a problem. By interface nitridation, generation of very fast interface states[13] and oxide hole traps[14] was indicated, and for $POCl_3$ annealing, generation of electron and hole traps in the oxide[15] was suggested.

It is more desirable if high-quality interface can be obtained without introducing foreign atoms into the interface and oxide. In recent years, ultrahigh-temperature oxidation (~ 1400 - 1600°C),[16] thin (~ 15 nm) oxidation with rapid cooling (> 600°C/min),[17] and post-oxidation argon (Ar) annealing[18] have been reported to be effective in reducing the interface states, without introduction of foreign atoms. In this study, we demonstrate that the $D_{IT}$ reduction by the post-oxidation Ar annealing is not the effect of "pure" thermal annealing but the effect of unintentional very-low-oxygen-partial-pressure ($p_{O2}$) annealing.

Samples employed in this study were n-type SiC (0001) MOS capacitors (donor concentration ($N_D$): ~ $10^{16}$ cm$^{-3}$). Oxides were formed by dry oxidation at 1300°C for 30 min followed by annealing in either pure Ar or Ar containing very small amount of oxygen ($O_2$) (0.001 - 0.1%; low-$p_{O2}$) at 1300 - 1500°C for 1 min. In the case of pure-Ar annealing, Ar was purified so that the concentration of contained oxygen was below 100 ppt. For low-$p_{O2}$ annealing, the partial pressure of oxygen was strictly controlled by supplying a gas directly from a gas cylinder containing the mixture gas of Ar and $O_2$. The low-$p_{O2}$ annealing was performed in an induction heating furnace with a fast cooling rate (> 600 °C/min), in order to minimize the additional oxidation during the cooling phase. The oxide thicknesses were about 51 - 58 nm and the gate electrodes were aluminum (Al) (diameter: 0.5 - 1 mm).

All of the measurements were conducted at room temperature.

Figure 1 shows the current density-electric field ($J$-$E$) characteristics of the prepared MOS capacitors. Here, the oxide electric field, $E_{OX}$, was estimated by $E_{OX} = V/t_{OX}$, where $V$ and $t_{OX}$ are the applied voltage and the oxide thickness, respectively. We see that the as-oxidized and low-$p_{O2}$-annealed samples exhibit typical $J$-$E$ characteristics with *Fowler-Nordheim* (*F-N*) tunneling current[2,19] at a sufficiently high oxide field (> 7 MVcm$^{-1}$). In contrast, a high leakage current is observed even at a very low field (< 0.2 MVcm$^{-1}$) in the case of pure-Ar-annealed sample.

In order to clarify the origin of the leakage current, secondary ion mass spectrometry (SIMS) measurements were performed. Figure 2 depicts the depth profiles of carbon concentration in the SiC/SiO$_2$ samples acquired by SIMS. After annealing in pure Ar, a high concentration of carbon atoms (> 10$^{20}$ cm$^{-3}$) is detected in the oxide, as reported in Ref. 6. Note that such a pure Ar ambient cannot be realized simply by introducing Ar immediately after the oxidation process, since residual oxygen remains in the oxidation furnace. In our case, we excluded the effect of oxygen by performing the Ar annealing in a leak-tight resistive heating furnace which is different from the oxidation furnace. In the case of low-$p_{O2}$-annealed sample, the carbon concentration in the oxide is close to the detection limit (~ 10$^{18}$ cm$^{-3}$). Thus, carbon atoms are ejected from the interface during the annealing, and they remain in the oxide in the case of pure Ar annealing, which leads to the severe degradation of oxide dielectric property (Fig.1). In the case of low-$p_{O2}$ annealing, slight oxygen (0.001 - 0.1%) removes the ejected carbon atoms by oxidizing them into gas species such as CO or $CO_2$, leading to the suppression of the leakage current (Fig.1).

Figures 3 (a) and (b) show the capacitance-voltage (*C-V*) characteristics of the MOS capacitors; annealed in the temperature range of 1300 - 1500°C in (a) $O_2$ (0.001%) and in (b) $O_2$ (0.1%). Both the frequency dispersion and *C-V* stretch-out are reduced by the low-$p_{O2}$ annealing, which is indicative of reduction of $D_{IT}$ near $E_C$. The effective fixed charge densities estimated from the flat-band voltage shift were $2.4 \times 10^{12}$ cm$^{-2}$ (negative) and $1.2 \times 10^{12}$ cm$^{-2}$ (positive) in the as-oxidized sample and the sample annealed in $O_2$ (0.001%) at 1500°C, respectively.

Energy distributions of $D_{IT}$ extracted by a high (1 MHz)-low method[19] are compared in Fig.4. In the case of annealing in $O_2$ (0.001%), the $D_{IT}$ is effectively reduced by increasing the temperature, and takes its minimum values (e.g. $2.4 \times 10^{12}$ eV$^{-1}$cm$^{-2}$ at $E_C - 0.2$ eV) after annealing at 1500°C. For $O_2$ (0.1%) annealing, in contrast, the $D_{IT}$ at $E_C - 0.2$ eV increases from $4.9 \times 10^{12}$ eV$^{-1}$cm$^{-2}$ to $6.3 \times 10^{12}$ eV$^{-1}$cm$^{-2}$ by increasing the temperature from 1300°C to 1500°C. The oxide thicknesses of the as-oxidized and low-$p_{O2}$-annealed samples determined by spectroscopic ellipsometry are summarized in Fig.5. We see that the oxide thickness hardly changes (< 1 nm) with the annealing in $O_2$ (0.001%), whereas, the thickness increases by about 6 nm with annealing in $O_2$ (0.1%) at 1500°C. Such results indicate that the $D_{IT}$ is determined by the balance of the removal and creation of the interface defects during the low-$p_{O2}$ annealing, and that it is important to avoid excessive oxidation of SiC during the low-$p_{O2}$ annealing to suppress additional defect generation. Note that, in SiC MOS systems, it is known that the incorporation of impurities, such as boron (B),[20] P,[12] and sodium (Na),[21] in a high concentration of about $10^{20}$ - $10^{21}$ cm$^{-3}$ leads to remarkable reduction of $D_{IT}$.[12,20,21] From SIMS measurements, we confirmed that the concentration of B, P, and Na atoms near the SiC/SiO$_2$ interface is at least below $2\times10^{16}$ cm$^{-3}$ after the low-$p_{O2}$ annealing, indicating that the observed $D_{IT}$ reduction by the low-$p_{O2}$ annealing (Fig.4) is not due to the impurity contamination.

We indicated that a high density of positive fixed charge ($1.2\times10^{12}$ cm$^{-2}$) resides in the sample annealed in $O_2$ (0.001%) at 1500°C. It should be noted that, it is difficult to estimate the real positive fixed charge density simply from the flat-band voltage shift in the case of as-oxidized sample, since electrons trapped at the acceptor-like interface states act as "effective" negative charge and compensates the positive charge. Thus, the positive charge may even reside in the as-oxidized sample and may become apparent by the low-$p_{O2}$ annealing owing to the reduction of acceptor-like interface states (Fig.4).

Here, we discuss the possible atomistic configurations of the major interface defects in SiC/SiO$_2$ systems. It has widely been believed that the carbon byproducts at (or near) the interface are the origin of interface states in SiC MOS structures,[2,6,16,17,22-27] since carbon is one of the host atoms of SiC. We confirmed that a high concentration of carbon atoms is ejected from the interface by the pure Ar annealing (Fig.2), which also suggests that the interface defects are related to carbon species. A result of density-functional calculations indicates that, among the various forms of carbon atoms that are frequently observed at a SiC/SiO$_2$ interface during molecular dynamics (MD) simulations[28], ethylene-like structure (SiO>C=C<SiO) creates defect levels near the $E_C$ of SiC.[23] Si$_2$>C=C<Si$_2$ defect[24-27] and Si$_2$>C=O defect[24] are also possible candidates, since they also create defect levels near the $E_C$. It is also suggested that oxygen helps the dissociation of interface carbon defects by reducing the energy of the structure after the dissociation by terminating the Si dangling bonds at the interface.[23] Thus, the low-$p_{O2}$ annealing may reduce the $D_{IT}$ near $E_C$ (Fig.4) by dissociating the carbon defects while preventing the additional generation of carbon defects caused by excessive oxidation of SiC.

In conclusion, we found that low-$p_{O2}$ annealing is effective in reducing the $D_{IT}$ at a SiC (0001)/SiO$_2$ interface without introduction of foreign atoms. For annealing in O$_2$ (0.001%), the $D_{IT}$ decreased by increasing the temperature up to 1500°C, whereas, for $O_2$ (0.1%) annealing, the $D_{IT}$ increased by increasing the temperature from 1300°C to 1500°C. The oxide thickness hardly changed (< 1 nm) with the annealing in $O_2$ (0.001%), whereas the thickness increased by about 6 nm with annealing in $O_2$ (0.1%) at 1500°C. Thus, during the low-$p_{O2}$ annealing, it is of importance to remove the interface defects by oxidizing them, while preventing the excessive oxidation of SiC to minimize additional defect creation. In the case of pure Ar annealing, carbon atoms are ejected from the interface, and they remain in the oxide, which leads to the severe degradation of oxide dielectric property. In low-$p_{O2}$ annealing, however, slight oxygen (0.001 - 0.1%) removes these carbon atoms by oxidizing them into gas species such as CO or $CO_2$, leading to the suppression of the leakage current.

This work was supported in part by the JSPS KAKENHI (Grant Number 15J04823) and the Super Cluster Program from the Japan Science and Technology Agency.

1) B. J. Baliga, *IEEE Electron Device Lett.* **10**, 455 (1989).

2) T. Kimoto and J. A. Cooper, *Fundamentals of Silicon Carbide Technology* (John Wiley & Sons Singapore, 2014).

3) N. S. Saks, S. S. Mani, A. K. Agarwal, *Appl. Phys. Lett.* **76**, 2250 (2000).

4) H. Yoshioka, J. Senzaki, A. Shimozato, Y. Tanaka, and H. Okumura, *AIP Advances* **5**, 017109 (2015).

5) T. Hatakeyama, Y. Kiuchi, M. Sometani, S. Harada, D. Okamoto, H. Yano, Y. Yonezawa, and H. Okumura, *Appl. Phys. Express* **10**, 046601 (2017).

6) T. Kobayashi and T. Kimoto, *Appl. Phys. Lett.* **111**, 062101 (2017).

7) S. Dhar, S. Haney, L. Cheng, S.-R. Ryu, A. K. Agarwal, L. C. Yu, and K. P. Cheung, *J. Appl. Phys.* **108**, 054509 (2010).

8) G. Y. Chung, C. C. Tin, J. R. Williams, K. McDonald, M. Di Ventra, S. T. Pantelides, L. C. Feldman, and R. A. Weller, *Appl. Phys. Lett.* **76**, 1713 (2000).

9) P. Jamet, S. Dimitrijev, and P. Tanner, *J. Appl. Phys.* **90**, 5058 (2001).

10) L. Lipkin, M. Das, G. Chung, J. Williams, N. Saks, and J. Palmour, *Mater. Sci. Forum* **383-393**, 985 (2002).

11) T. Kimoto, Y. Kanzaki, M. Noborio, H. Kawano, and H. Matsunami, *Jpn. J. Appl. Phys.* **44**, 1213

(2005).

12) D. Okamoto, H. Yano, T. Hatayama, and T. Fuyuki, *Appl. Phys. Lett.* **96**, 203508 (2010).

13) H. Yoshioka, T. Nakamura, and T. Kimoto, *J. Appl. Phys.* **112**, 024520 (2012).

14) Y. Katsu, T. Hosoi, Y. Nanen, T. Kimoto, T. Shimura, and H. Watanabe, *Mater. Sci. Forum* **858**, 599 (2016).

15) H. Yano, N. Kanafuji, A. Osawa, T. Hatayama, and T. Fuyuki, *IEEE Trans. Electron Devices* **62**, 324 (2015).

16) T. Hosoi, D. Nagai, M. Sometani, Y. Katsu, H. Takeda, T. Shimura, M. Takei, and H. Watanabe, *Appl. Phys. Lett.* **109**, 182114 (2016).

17) R. H. Kikuchi and K. Kita, *Appl. Phys. Lett.* **105**, 032106 (2014).

18) T. Kobayashi, J. Suda, and T. Kimoto, *AIP Advances* **7**, 045008 (2017).

19) S. M. Sze and Kwog K. Ng, *Physics of Semiconductor Devices, 3rd edition* (John Wiley & Sons, Inc., Hoboken, New Jersey, 2007).

20) D. Okamoto, M. Sometani, S. Harada, R. Kosugi, Y. Yonezawa, and H. Yano, *IEEE Trans. Electron Devices* **35**, 1176 (2014).

21) G. Gudjónsson, H. Ö. Ólafsson, F. Allerstam, P.-Å. Nilsson, E. Ö. Sveinbjörnsson, H. Zirath,

T. Rödle, and R. Jos, *IEEE Electron Device Lett.* **26**, 96 (2005).

22) V. V. Afanas'ev, M. Bassler, G. Pensl, and M. Schulz, *Phys. Status Solidi A* **162**, 321 (1997).

23) Y. Matsushita and A. Oshiyama, arXiv:1612.00189 (2016).

24) T. Kaneko, N. Tajima, T. Yamasaki, J. Nara, T. Schimizu, K. Kato, and T. Ohno, *Appl. Phys. Express* **11**, 011302 (2018).

25) F. Devynck, A. Alkauskas, P. Broqvist, and A. Pasquarello, *Phys. Rev. B* **83**, 195319 (2011).

26) P. Deák, J. M. Knaup, T. Hornos, C. Thill, A. Gali, and T. Frauenheim, *J. Phys. D* **40**, 6242 (2007).

27) Y. Matsushita and A. Oshiyama, *Jpn. J. Appl. Phys.* **57**, 125701 (2018).

28) R. Car, and M. Parrinello, *Phys. Rev. Lett.* **55**, 2471 (1985).

Fig.1: Current density-electric field (*J-E*) characteristics of as-oxidized and low-$p_{O2}$-annealed SiC MOS structures.

Fig.2: Depth profiles of carbon concentration in SiC/SiO$_2$ samples acquired by secondary ion mass spectrometry (SIMS). Note that the profiles a few nanometers from the SiO$_2$ surface are unreliable owing to the initial non-steady state in dynamic SIMS.

Fig.3: Capacitance-voltage (*C-V*) characteristics of as-oxidized and low-$p_{O2}$-annealed SiC MOS structures; (a) effect of annealing in O$_2$ (0.001%) and (b) in O$_2$ (0.1%).

Fig.4: Energy distributions of interface state density ($D_{IT}$) in as-oxidized and low-$p_{O2}$-annealed SiC MOS structures extracted by a high (1 MHz)-low method.

Fig.5: Oxide thicknesses of as-oxidized and low-$p_{O2}$-annealed SiC/SiO$_2$ samples determined by spectroscopic ellipsometry.

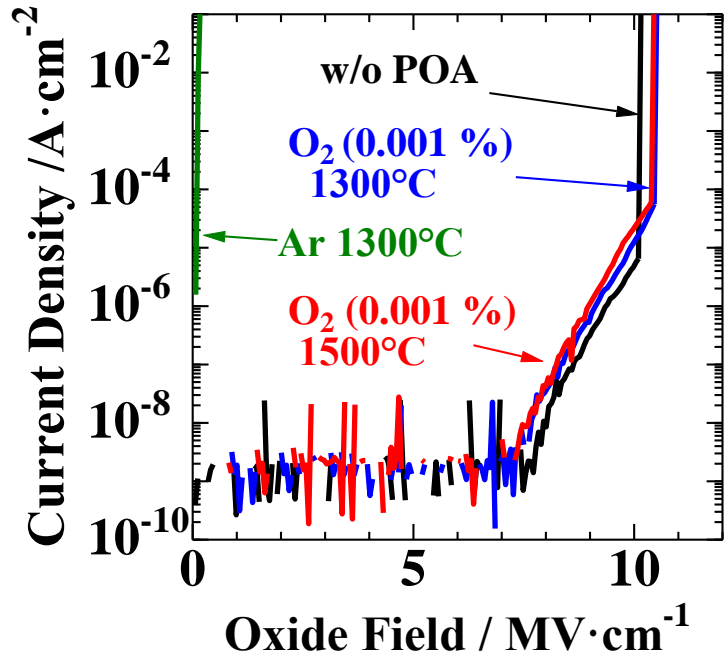

Fig.1

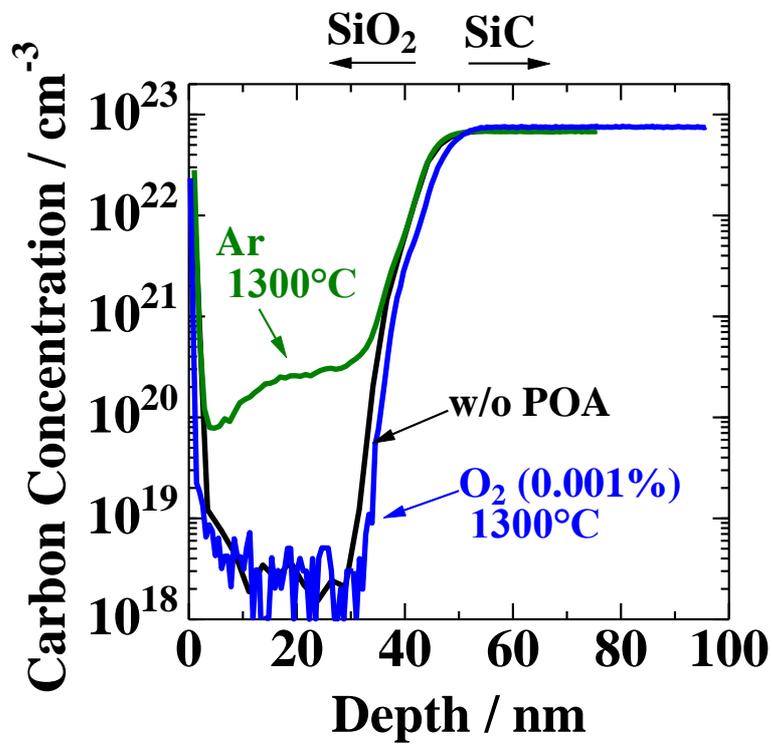

Fig.2

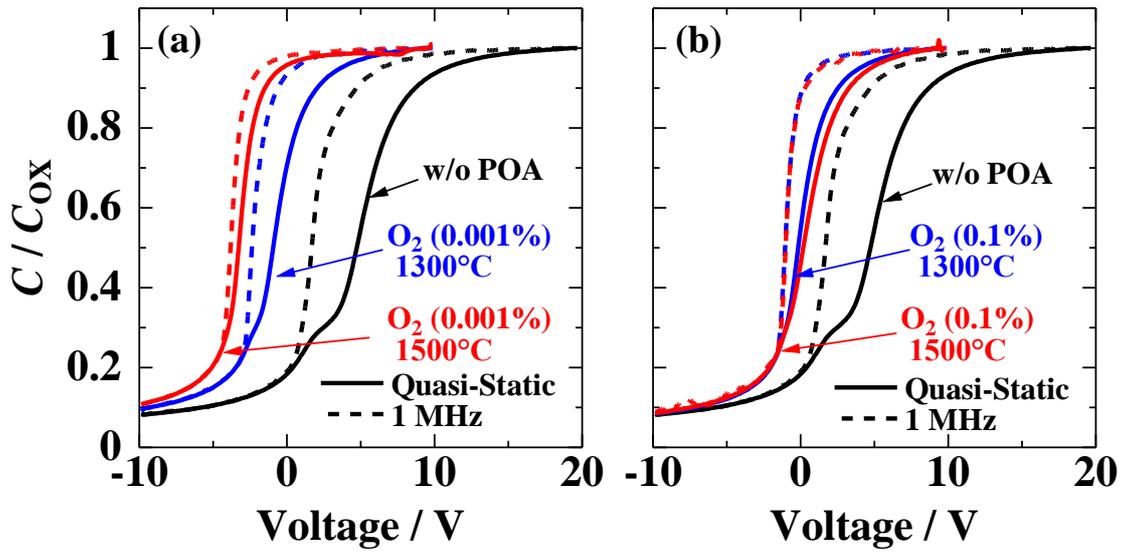

Fig.3

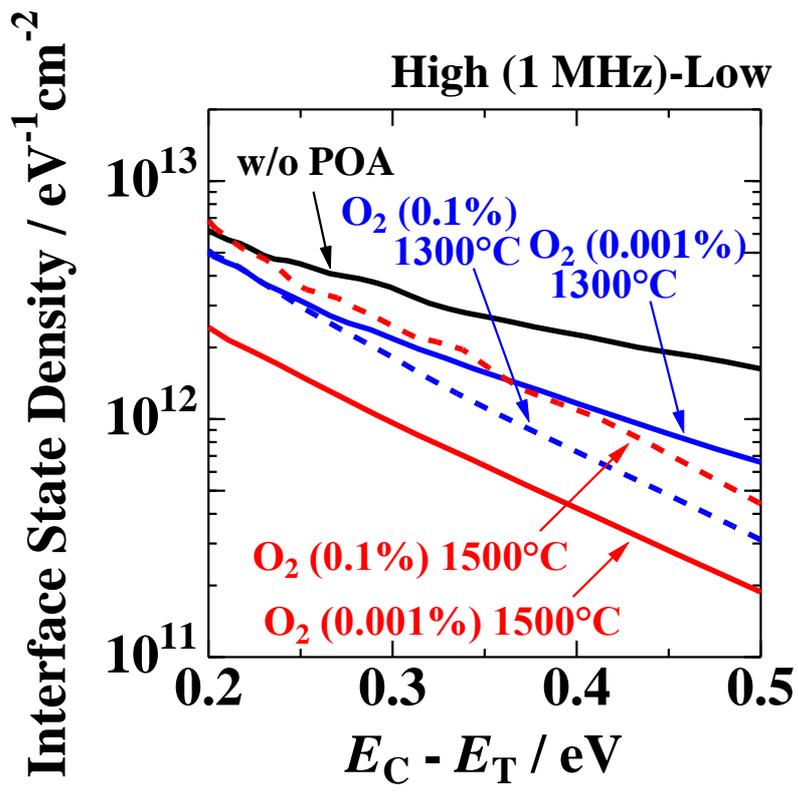

Fig.4

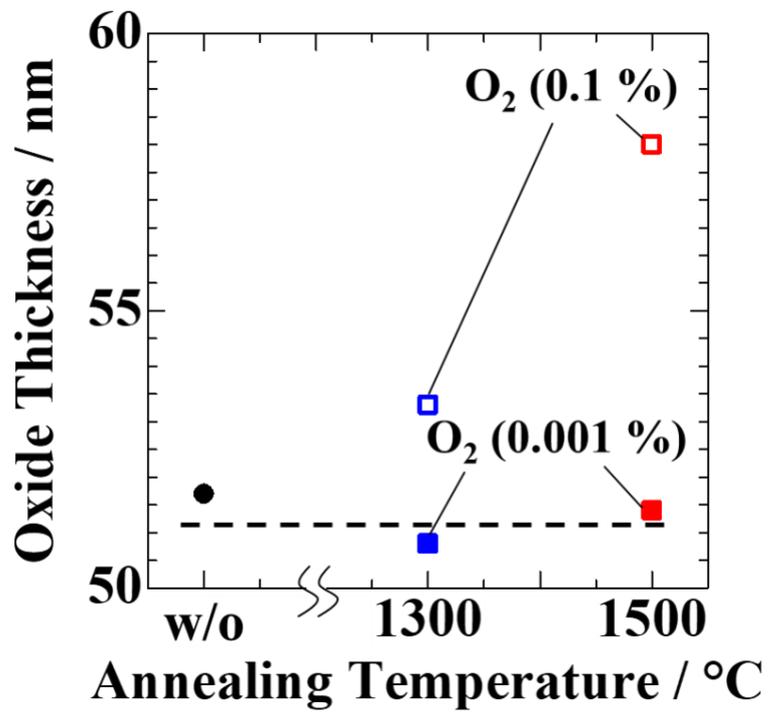

Fig.5